# The Impact of Generative Artificial Intelligence on Ideation and the performance of Innovation Teams (Preprint)

24.08.2024

Gindert Michael* M. Sc., *University Regensburg,*
Müller Marvin Lutz, M. Sc. with Honors, *University Regensburg*

## Abstract

This study investigates the impact of Generative Artificial Intelligence (GenAI) on the dynamics and performance of innovation teams during the idea generation phase of the innovation process. Utilizing a custom AI-augmented ideation tool, the study applies the Knowledge Spillover Theory of Entrepreneurship to understand the effects of AI on knowledge spillover, generation and application. Through a framed field experiment with participants divided into experimental and control groups, findings indicate that AI-augmented teams generated higher quality ideas in less time. GenAI application led to improved efficiency, knowledge exchange, increased satisfaction and engagement as well as enhanced idea diversity. These results highlight the transformative role of the field of AI within the innovation management domain and shows that GenAI has a positive impact on important elements of the Knowledge Spillover Theory of Entrepeneurship, emphasizing its potential impact on innovation, entrepreneurship, and economic growth. Future research should further explore the dynamic interaction between GenAI and creative processes.

# 1 Introduction

The development of artificial intelligence (AI) systems is advancing at a much faster pace than experts worldwide had anticipated. Creativity and novel ideas are the driving forces behind innovation [1] and, as such, are central to economic growth. [2], [3] Therefore, the impact of AI on these domains is of particular interest. Recent studies underscore the creativity exhibited by GenAI, demonstrating higher continuity, flexibility, and originality in idea generation compared to human participants [4]. Research clearly indicates that GenAI can significantly enhance the quality of idea generation [5]. GenAI, particularly using Large Language Models (LLMs), has therefore the potential to profoundly transform innovation management, impacting every phase of the innovation process [6], [7], [8], [9]. Given recent advancements, GenAI may hold the greatest potential in the ideation phase of the innovation process [10] The collaboration between human and artificial intelligence is increasingly recognized as a critical element in enhancing decision-making and creativity within organizations [11], [12], [13], [14] Creativity Support Systems (CSS) have been developed to improve the flow of ideas, novelty, and diversity of individually generated concepts, which are inherently constrained by the user's personal knowledge. These tools allow users to explore a broader spectrum of creative possibilities, facilitating the development of ideas that might not have emerged through intuition alone [15]. The integration of GenAI, particularly LLMs, has the potential to make these systems even more effective.

Despite the burgeoning interest and research in GenAI, the integration of AI tools into human-centric innovation processes, particularly in team-based settings, remains underexplored. Recent studies, such as the comparative analysis of human-generated and ChatGPT-generated ideas, have focused on evaluating AI-generated ideas in isolation [5]. However, the specific impact of AI-supported idea generation on team dynamics and the quality, speed, and diversity of ideas generated collaboratively by humans with AI assistance has not been adequately addressed. This gap in the literature presents a significant opportunity to explore the potential synergies between AI and human creativity. Specifically, there is a need to understand how AI-augmented ideation tools influence the overall ideation outcome in collaborative settings. Do AI-supported tools enhance the quality and efficiency of ideas generated by human teams? Can these tools accelerate the ideation process and increase the diversity of ideas produced? These questions are critical as organizations increasingly look to leverage AI to boost their innovation capabilities. The current study aims to rigorously investigate the effects of AI-supported idea generation on team-based innovation, focusing on the quality, speed, and diversity of ideas.

Utilizing an AI augmented ideation tool based on GPT-4 Turbo, this research compares the performance of teams using AI tools with those that do not. Participants, organized into three-person teams, tackle identical tasks, with expert evaluations assessing the generated ideas. The objective is to analyze and quantify the impact of AI on the ideation processes within team-based innovation settings, offering empirical insights into AI's role in shaping team creativity and innovation outcomes. The scope centers on evaluating these impacts through a controlled experimental design, applying descriptive and inductive statistical methods to understand the interplay between AI and human creativity. The study seeks to provide valuable contributions by identifying how AI support influences ideation within innovation teams, highlighting both the potential benefits and limitations of AI augmentation in real-world innovation contexts.

The Knowledge Spillover Theory of Entrepreneurship (KSTE) elucidates the dynamics of knowledge generation, highlighting the pivotal roles played by established firms, universities, and public research institutions in research and development activities [16], [17]. A central focus of KSTE research is the identification of different types of knowledge spillover and the mechanisms through which they occur [18]. There are various potential ways in which Generative AI (GenAI) might influence these mechanisms. For example, GenAI may indirectly impact entrepreneurs during their creative processes or directly affect key aspects of KSTE, such as altering the knowledge filter and enhancing absorptive capacity. Building upon the foundational insights of field experiment design provided by Harrison and List (2004) [19], this study is structured into two distinct phases. Participants are randomly assigned to either experimental or control groups and are organized into three-person teams. These teams are tasked with solving two identical problems. The experimental group utilizes a custom-built AI-assisted idea generation tool, based on OpenAI’s GPT-4 Turbo, while the control group does not receive any AI support. Subsequently, experts—blinded to the group assignment—evaluate the generated ideas based on the quality dimensions outlined by Dean et al. (2006) [20]. This quantitative study adopts a deductive research approach and applies both descriptive and inferential statistical methods to analyze the data. The primary objective is to assess the impact of AI-assisted tools on the quality, speed, and diversity of idea generation within team-based innovation processes. The findings of this study aim to contribute to a deeper understanding of the interplay between AI and human creativity, with implications for enhancing innovation outcomes in organizational settings.

This paper investigates the impact of recent advancements in Artificial Intelligence, particularly Generative AI and LLMs, the knowledge spillover process during the ideation phase in team-based innovation settings. The data indicates that the experimental group, which utilized

GenAI, significantly outperformed the control group across multiple dimensions such as originality, clarity, and completeness of ideas. This superiority was evident in tasks related to healthcare and automotive industry innovations. Despite some variability in applicability and effectiveness, the use of GenAI demonstrated a robust positive impact on the overall quality and efficiency of idea generation, with the experimental group achieving higher satisfaction and faster completion times. This work first explores the theoretical concepts of creativity and innovation. The work explores key concepts such as the impact of AI, specifically generative AI, on creativity, idea generation, and innovation team performance, structured through a detailed literature review on creativity, innovation and artificial intelligence, an empirical study involving a controlled field experiment, and subsequent data analysis to quantify AI's influence on ideation and innovation processes.

# 2 Creativity and the dynamics of Innovation

## 2.1 Creativity as a driver of Innovation

Creativity is a complex socio-psychological phenomenon that drives innovation through the generation of novel and useful ideas [21], [22], [23]. The creative process encompasses problem identification, idea generation, and implementation [24]. Imagination seems to play an important role in fostering creativity - a concept emphasized by Einstein's (1929) statement ‘Imagination is more important than knowledge’[25] and supported by various creativity theories [26], [27], [28], [29]. Creativity can be defined as a process that arises from a perception of the environment that recognizes a certain imbalance and leads to a productive activity that challenges patterned thought processes and norms and produces something new in the form of a physical object or even a mental or emotional construct [30]. Creativity emerges from the interplay between reproduction (knowledge and past experiences) and projection (imagination), triggered by perceived discrepancies in one's environment [31]. The transition from creativity to innovation involves a spectrum of contributions, ranging from replication to integration [32] while balancing originality and practicality [33]. Cropley (2006) describes creativity as the engine of innovation [1], [34]. The Componential Model of Creativity and Innovation emphasizes the interplay of individual and organizational factors in the innovation process [35]. Idea generation is a key process in organizational innovation, where cognitive flexibility is crucial for the development and refinement of different thought elements into actionable concepts [36], [37], [38]. Creativity involves combinatorial and exploratory approaches that require a comprehensive knowledge base and the ability to manipulate existing knowledge to develop meaningful and valuable ideas [39], [40]. In the field of computational creativity, the focus is on replicating or enhancing human creativity, with bisociation playing a key role in linking disparate domains to generate novel and useful ideas [40], [41], [42]. Inspirational stimuli are crucial in this process, as they foster efficient ideation and lead to novel, unique, and ultimately practical solutions. The impact of these stimuli on creativity is significant, often surpassing individual problem-solving abilities [43], [44], [45], [46]. Research on group creativity shows that fear of sharing ideas and production blockages can hinder brainstorming, reducing both group performance and motivation [47], [48], [49]. Positive outcomes are more likely when groups are exposed to diverse stimuli and ideas and are motivated by increased responsibility and competition [50], [51]. Electronic brainstorming, particularly when groups are given more time, can reach the idea generation level of traditional groups and overcome the blocking effect [51]. Typically, more ideas are produced in electronic brainstorming than in traditional brainstorming, as the production blocking effect is reduced, and effectiveness increases directly with group size [50], [52], [53], [54].

## 2.2 Dynamics of Innovation

When exploring the intersection of entrepreneurship, economic growth, and technology, foundational theories provide valuable insights. The Solow model demonstrates technological progress as main driver for economic growth [55], Schumpeter’s concept of creative destruction emphasizes the essential role of innovation and entrepreneurship in fostering long-term economic development and reallocating resources [56]. Expanding on these ideas, Romer’s endogenous growth model identifies knowledge and human capital as key drivers of sustained economic growth, laying the groundwork for the Knowledge Spillover Theory of Entrepreneurship (KSTE) [57].

KSTE proposes that the spillover of knowledge, particularly in entrepreneurial contexts, is crucial for economic advancement [58]. The theory suggests that not all newly created knowledge is commercialized, creating entrepreneurial opportunities [16], [17] as uncommercialized knowledge spills over to economic actors, influencing resource allocation[16]. Much like the advent of the internet, it seems that recent advances in the AI technology trend, particularly GenAI, could significantly enhance societal growth by making knowledge more accessible to individuals and organizations [59], [60], [61]. GenAI, especially in the form of LLMs, could provide direct access to rich and diverse information, revolutionizing the way entrepreneurs and researchers interact and apply knowledge. The novelty of recent advances in the field of AI means that its role in knowledge spillovers and entrepreneurial innovation is still underexplored and requires further research.

In Romer (1990) and Lucas's (1993) endogenous growth models, knowledge is recognized as a key driver of growth that naturally spills over to others [62], [63]. Later theories introduced the idea of a 'knowledge filter,' which hinders the automatic spread of information [64], [65], [66], [67]. These filters—organizational, commercial, and economic—create barriers along the path from startup to established organization [68]. Arrow (1962) distinguishes between new, unused knowledge and economically valuable knowledge[69], with KSTE focusing on how knowledge spillovers occur, how they are captured, and their application in entrepreneurship. While universities, research institutions, and research-intensive industries are vital knowledge sources, much created knowledge remains uncommercialized. Braunerhjelm et al. (2010) highlight that only 1-2% of knowledge reaches the market, and only 25% of inventions become patents, leaving the majority as unutilized knowledge [70]. Entrepreneurs play a critical role in leveraging this knowledge, driving economic growth by transforming tacit knowledge into explicit knowledge through processes like Socialization, Externalization, Combination, and Internalization (SECI). This transformation is facilitated in collaborative environments where knowledge exchange occurs physically or virtually, supporting the development of entrepreneurial ecosystems [71], [72].

Absorptive capacity, both cognitive and technological, also plays a crucial role in this process [73], enabling the assimilation and application of knowledge, often requiring investments in R&D and industry-specific technologies [74]. Absorptive capacity refers to the capability to recognize, internalize, modify, and apply external expertise, findings, and methodologies [75]. Entrepreneurs' capacity to leverage knowledge spillovers and drive innovation through startups hinges on their individual absorptive capacity [59], [76], which is essential for transforming new knowledge into commercial applications and fostering regional economic growth [2], [73]. The decision-making process in newly established companies, significantly influenced by the founders' cognitive and technical skills, as well as their knowledge, impacts the company's development and success in the competitive market [73], [77]. As innovators, entrepreneurs face the challenge of navigating the complex knowledge filter and strategically leveraging their absorptive capacity to effectively utilize knowledge spillover [71], [76]. A deeper understanding of how knowledge and the resulting innovations can be implemented is provided by innovation management.

Innovation management has evolved significantly, shifting from a closed, linear process to a more open, iterative, and collaborative approach, driven by the widespread dissemination of knowledge and rapid socio-technological changes[78], [79], [80]. Within the innovation process [81], the ideation phase is crucial as it serves as the foundation for innovation, enabling the exploration of diverse possibilities and the identification of opportunities that can lead to groundbreaking advancements and competitive advantage. In assessing the quality of ideas at this stage, Dean et al. (2006) propose a framework focusing on novelty, feasibility, relevance and specificity for assessing the potential of innovations [20]. This approach ensures ideas are innovative, practical, impactful, and clearly communicated for execution. Ideas can also be categorized into the Abernathy and Clark Innovations Matrix (1985) that categorizes innovations based on their impact on existing technological competencies and market relationships. It distinguishes four types of innovation: incremental, architectural, niche and revolutionary [82].

# 3 Generative Artificial Intelligence and Innovation

## 3.1 GenAIs role within the innovation process model

Natural Language Processing (NLP) has evolved from basic text retrieval to becoming integral to human-computer interaction [83], [84]. Transformer-based models, especially those utilizing self-attention mechanisms [85], have significantly advanced text processing, sparking interest in their application within innovation [84], [86], [87]. Studies highlight GPT-4's potential in creative tasks, emphasizing AI's growing role in routine and complex problem-solving [4], aligning with the early visions of AI pioneers [88]. Generative AI is now reshaping creativity across industries[89], [90], driving hybrid human-AI operating models and new roles in innovation management[91], [92], [93].

As companies adapt their innovation strategies to leverage this technological progress, generative AI is gaining attention for its potential to complement or even replace human tasks within organizational innovation processes [6], [10], [94]. High research and development investments do not always equate to significant innovations, as a company's ecosystem[95], [96], [97], access to knowledge [98], [99], and political support are crucial [100], [101], [102], [103]. The integration of GenAI into innovation processes is increasingly emphasized due to its potential to augment or replace human tasks [104], [105], particularly in information processing and decision-making within innovation management [106], [107]. Generative AI impacts each stage of the innovation process [108], [109], [110], [111] aiding in setting objectives, rationalizing data processing, and generating valuable insights [106], [107]. GenAI has the potential to significantly enhance creativity by overcoming human limitations in idea generation, particularly in four key areas: first, by surpassing constraints in information processing to develop (1) and generate (2) ideas; and second, by overcoming local search routines to generate (3) and develop (4) ideas [10]. GenAI's capacity to process large datasets facilitates advancements across various domains, from material discovery and pharmaceutical research and development to organizational process innovations, which have already yielded substantial economic benefits [112], [113], [114]. Examples like the Graph Networks for Materials Exploration (GNoME) demonstrate how generative AI can accelerate materials discovery and innovation in diverse domains [115].

## 3.2 Large Language Models and knowledge

LLMs possess the ability to store and retrieve diverse knowledge from their training data, which can be accessed through fine-tuning and prompt engineering, positioning them as unsupervised knowledge retrieval systems [116], [117], [118], [119], [120]. LLMs are therefore increasingly valuable for knowledge spillovers and entrepreneurial activities. While traditional knowledge extraction from unstructured data was labor-intensive[121], transformer models like GPT-3 and GPT-4 have greatly improved efficiency, benefiting innovation teams [122], [123].

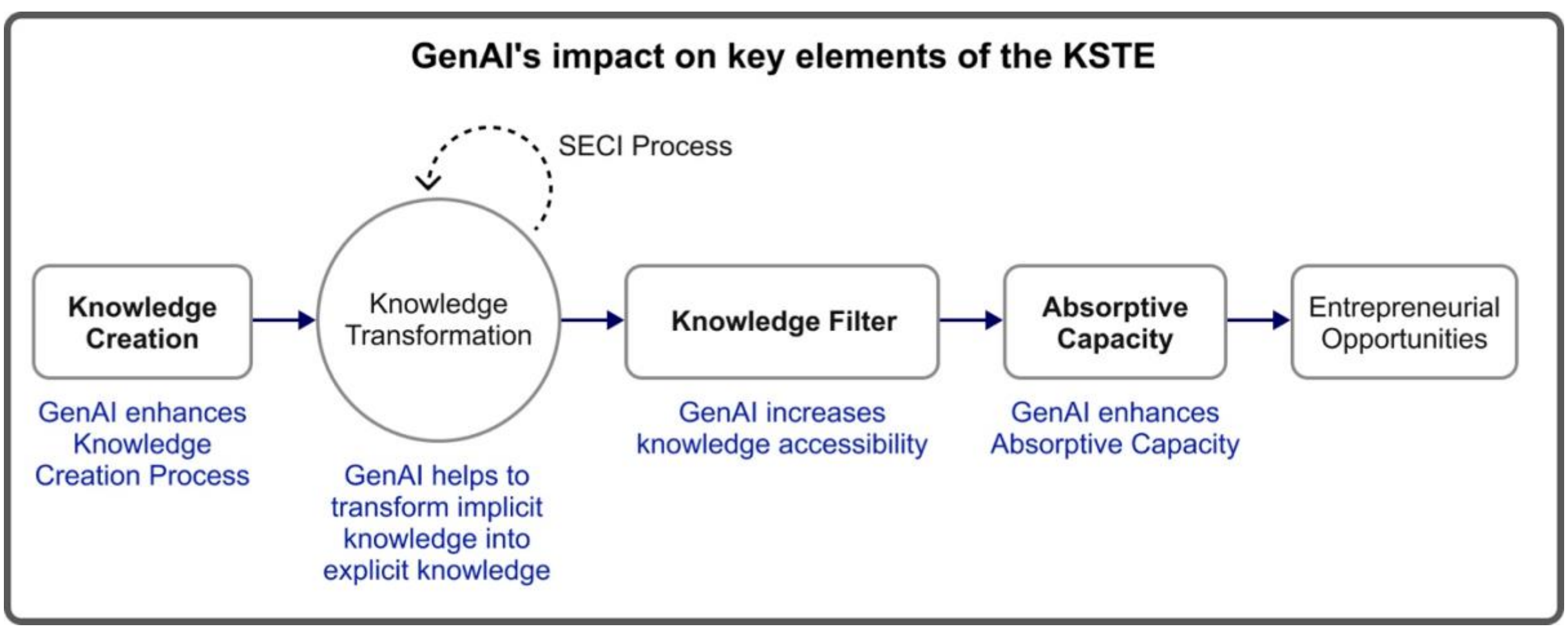

*Figure 1: Interaction interfaces of GenAI on central aspects of KSTE*

GenAI significantly enhances knowledge management by facilitating knowledge spillover and creation, driving innovation [124], [125], [126]. As shown in Figure 1, LLMs could change important elements of knowledge spillovers by converting implicit knowledge into explicit knowledge and analyzing complex relationships within training data [127], [128]. LLMs also have the potential to transform traditional knowledge filters by making knowledge more accessible. By improving knowledge creation and availability, LLMs might enhance absorptive capacity by contextualizing information, enabling more effective knowledge spillovers among various actors. Furthermore, LLMs accelerate information processing, increasing the speed and efficiency of knowledge dissemination. Research underscores the importance of both geographical and virtual platforms in entrepreneurial ecosystems for accessing knowledge spillovers [129], [130], with proximity to knowledge sources being a key factor [59], [60]. LLMs fundamentally alter this dynamic by improving online access to knowledge. Techniques like hierarchical transfer and prompt engineering, when applied to LLMs, support human knowledge work by encoding vast information, which can be used to create knowledge graphs for explicit knowledge generation [131], [132], [133], [134]. Organizations can enhance absorptive capacity through investments in research and development, as well as information and communication technologies (ICTs), including LLMs [74], [135]. Start-ups leveraging these capacities can use ICT tools to capture and process explicit knowledge more effectively [76], [135], [136], [137], [138].

## 3.3 Application of generative AI in ideation

Creativity at the individual level is rooted in the integration of primary resources such as knowledge, the competence for innovative synthesis of these resources, and intrinsic motivation as a driving force [31]. The ideation process is characterized by the application and synthesis of knowledge, stimulating cognitive processes [139], [140]. Creativity involves both combinatorial and exploratory approaches, requiring a broad knowledge base and the ability to manipulate this knowledge to generate valuable ideas [39], [40]. Inspiring stimuli are crucial in this process, leading to novel, unique, and practical solutions [43], [44], [45], [46]. The effectiveness of current LLMs is most evident in the ideation phase of the innovation process, where collaborative environments enhance knowledge spillover [71], [141]. GenAI can significantly boost creativity by overcoming human limitations in information processing and local search routines during idea generation [10]. GenAI's specific applications in problem exploration involve utilizing embedded knowledge within LLMs to identify innovation opportunities [142], a task previously limited by the difficulty of extracting knowledge from unstructured data [121]. In exploring solution spaces, GenAI models excel in ideation [143] and uncovering new technological possibilities [144], [145], significantly expanding both creative problem and solution spaces [145]. GenAI serves as a crucial catalyst for creativity in innovation, helping to overcome cognitive limitations and fostering an environment conducive to innovation. Creativity support systems (CSS), particularly those augmented by GenAI, are designed to enhance the knowledge flow, novelty, and diversity of ideas [15]. This work focuses on a GenAI-augmented CSS, specifically a GPT-4-supported system, developed for further research.

*Table 1: Experimental hypotheses*

| # | Hypothese | Reasoning | KPI |
|---|---|---|---|
| **H1** | The use of LLMs during the idea generation process will lead to greater knowledge spillover and thus to a higher quality (according to Dean et al. (2006)) of ideas compared to traditional idea generation techniques (without AI). | LLMs can process large amounts of information, recognise patterns and gain new insights, which can improve the idea generation process. | Evaluation criteria according to Dean et al., 2006 |
| **H2** | The use of LLMs in ideation will increase the speed of the ideation process in team-based innovation environments. | LLMs can quickly develop many ideas, which can speed up the ideation process. | Time measurement until the respective idea is found |
| **H3** | The use of LLMs will increase the efficiency of the idea generation process in team-based innovation environments. | LLMs can quickly develop many ideas and be elaborated into extensive ideas by human idea generators | Idea quality and time measurement for idea generation |
| **H4** | The use of LLMs will contribute to a greater diversity (in terms of novelty and feasibility) of ideas in team-based ideation sessions. | LLMs can generate a wide variety of ideas, which can lead to a greater diversity of ideas in the ideation sessions. | Evaluation criteria: Originality, paradigm relevance, feasibility |
| **H5** | The use of GenAI-augmented ideation tools will increase the likelihood of teams developing revolutionary ideas (Abernathy and Clark, 1985). | AI tools can generate new and unexpected ideas that can lead to more radical innovations. | Number and distribution of revolutionary ideas |
| **H6** | The use of LLMs will increase team member satisfaction and engagement during the ideation process. | LLMs could make the ideation process more engaging and less stressful, increasing team satisfaction. | Post-experiment survey on team satisfaction and commitment. |

As shown in Table 1, this study examines six main hypotheses on LLMs' impact on idea generation. These were tested through a controlled field experiment using both quantitative and qualitative data analysis.

# 4 Methodology

The objective of quantitative research is to test hypotheses, confirm assumptions and theories, and identify cause-effect relationships. In the context of causal-analytical comparative research, the study examines the influence of independent variables on dependent variables, establishing causal relationships between these factors. This causal-comparative research design allows the researcher to investigate the interaction between independent variables and their impact on dependent variables [146]. Field experiments serve as a valuable tool in this research, particularly when exploring how context influences decision-making in individuals and groups. Field experiments are well-suited for investigating causal relationships, measuring treatment effects in real-world settings by manipulating one or more independent variables and assessing their impact on one or more dependent variables [19]. In investigating the influence of GenAI on ideation and the performance of innovation teams, the goal is to explore the causal relationship between the use of AI as an independent variable and the quality and efficiency of outcomes as dependent variables. Therefore, a controlled field experiment is appropriate for addressing this research question.

Following the recommendations of Harrison and List (2004), a controlled field experiment was designed, consisting of two phases [19]. In the first phase, participants are randomly assigned to either an experimental group or a control group. Teams of three individuals then work on two tasks: designing a product/service innovation in the healthcare sector for seniors and developing a business model innovation in the automotive industry considering advances in autonomous driving. Each team produces a concrete solution proposal for both tasks, with the control group receiving no AI support during ideation and the experimental group utilizing a specially developed GenAI-assisted CSS. In the project's second phase, an expert panel—consisting of an industry innovator, entrepreneur, and innovation researcher—evaluates the solution proposals anonymously, using the criteria established by Dean et al. (2006) [20]. Subsequently, all ideas are categorized using the Abernathy and Clark (1985) innovation matrix [82].

This study employs a quantitative methodology, focusing on data evaluation, causal relationships, and hypothesis testing, which are fundamental to this approach. By using Likert scales to measure idea quality, it generates numerical data, facilitating the analysis of correlations and differences among variables. The study addresses the key research question regarding the influence of large language model-based ideation tools on knowledge spillover in team-based innovation settings. Following a deductive approach, the research builds on the Knowledge Spillover Theory of Entrepreneurship, examining the impact of LLMs on ideation and drawing conclusions from hypothesis testing that enhance our understanding of GenAI's role in innovation processes.

## 4.1 Research design and approach

A custom-designed ideation tool, developed specifically for the study (Appendix 1), was employed to compare traditional and GenAI-augmented brainstorming techniques. Refined through extensive UX research with fifteen participants across three usability studies, this tool was tailored to user needs, offering an intuitive, chat-like interface for a seamless collaborative experience. Study participants logged in with unique codes to ensure proper group assignment, anonymity, and secure data storage.

When a team member posed a problem, the tool's advanced algorithm broke it down into components, converting them into detailed queries for the GPT-4 backend (Appendix 1). For each problem, the system generated four AI personas, each with specialized expertise, tailored to the problem at hand to maximize output quality. These personas engaged dynamically in brainstorming sessions, offering insights and feedback like human experts (Appendix 1). Teams could regenerate personas with refined problem descriptions in a separate tab, enabling a flexible brainstorming process.

The tool's backend architecture featured a database interface that recorded every user interaction, idea, and solution proposal in real-time, ensuring no loss of data and providing a complete record for analysis. For task management, the tool allowed teams to display, discuss, and input solutions directly within the interface, with solutions only saved once consensus was reached, promoting collaborative decision-making. Additionally, the tool employed an automated prompt sequencing system, continuously feeding the ongoing conversation transcript into the AI to prevent repetitive outputs and ensure each contribution was unique and contextually relevant.

In this experiment, a critical element of AI software development was controlling the model's 'temperature,' which dictates the level of randomness in AI responses. By setting the temperature to zero, we ensured deterministic outputs, enhancing the comparability of the teams' results. The tool combined advanced GenAI features with user-focused design and strong data management to enable an innovative approach to ideation while providing a controlled environment for comparing GenAI-supported and traditional brainstorming. Additionally, to ensure unbiased evaluation, ideas generated by both groups were anonymized and combined into a curated database, preventing any traceability to the original teams by the expert jury.

The methodology involved two groups of students from the University of Regensburg: an experimental group and a control group, each working in separate rooms. Participants, unaware of their group assignment, completed identical tasks using the same background information.

Only the experimental group had access to the GenAI augmented ideation tool without prior knowledge of it.

The participants were business, economics, and information systems students with coursework in 'International Management' and 'Strategic Management,' ensuring a relevant foundation for exploring knowledge spillover. Their limited exposure to 'Healthcare' and 'Automotive Innovation' helped reduce external variables. The control group was not informed about the experimental group's use of GenAI to prevent the John Henry effect [147].

The study involved 70 students in 24 teams, generating 48 unique ideas. A point-based incentive system included bonus points and a one-month ChatGPT Plus subscription for the best idea. Teams worked at separate stations, with icebreakers to address brainstorming challenges. The control group began brainstorming after a brief introduction, with their task completion time recorded. The experimental group had an introduction and Q&A session about the AI tool, followed by access to the GenAI-supported platform, which logged inputs, completion times, and individual contributions. Both groups followed a 45-minute maximum task duration but could submit earlier if completed.

For ethical reasons, all inputs were stored anonymously, with unique acronyms assigned to users. Participants provided informed consent and could withdraw from the study at any time.

## 4.2 Data Collection and Analytical Methodology

The study's primary research question explores the impact of ideation tools, enhanced by LLMs, on knowledge spillover during the ideation phase in team-based innovation settings. This question drives the investigation into how LLMs affect the ideation phase and the performance of innovation teams.

The study's objectives, using quantitative research methods, are as follows:

- Compare the effectiveness of LLM-supported ideation tools with traditional methods in enhancing knowledge spillover and idea quality.
- Investigate how LLM integration affects the speed of idea generation in innovation teams.
- Evaluate the impact of LLM-supported ideation on the efficiency of team-based ideation processes.Assess the diversity of ideas generated with LLM support, focusing on novelty, feasibility, and relevance.
- Analyze the influence of GenAI tools on categorizing innovations as incremental, architectural, niche, or disruptive.

- Measure the effects of GenAI-supported ideation on team members' satisfaction and engagement.

Data collection in the experimental group was facilitated through the specialized ideation tool, while the control group used Google Forms. An expert panel evaluated the quality of ideas using Microsoft Forms on a 7-point Likert scale, categorizing them using the Abernathy and Clark (1985) innovation matrix [82]. Variables such as idea quality based on Dean et al. (2006) [20], processing times, and the evaluation criteria were rigorously operationalized and validated through usability tests, pilot studies, and calibration exercises.

Descriptive statistics were employed to assess the number of ideas, the time required, and the innovation category, with particular attention to the distribution of scores for each quality measure. Inferential statistics, including t-tests and multivariate linear regressions with multiple dependent variables, were used to compare the mean quality scores per idea and to evaluate the impact of group classification on multifactorial idea quality. A bootstrapping procedure was additionally used to confirm the robustness of the results [148].

# 5 Results

For the analysis, the Phyton based Data Science platform Anaconda 3 and Jupyter Notebooks were utilized to ensure an efficient and effective evaluation of the data.

The analysis starts with a descriptive overview of the experimental data. Overlapping histograms and kernel density estimates illustrate the score distributions across evaluation dimensions. Figure 2 visually compares the performance trends of the control and experimental groups for both tasks.

In the first task, focused on product or service innovations in elderly healthcare, the experimental group using a GenAI tool during ideation clearly outperformed the control group across several key dimensions, including aggregate score, acceptance, feasibility, completeness, and clarity. Similarly, in the second task, which dealt with business model innovations in the automotive industry, the experimental group also surpassed the control group in aggregate score, originality, feasibility, completeness, and clarity. The consistent results across both tasks justified combining the datasets to strengthen the analysis.

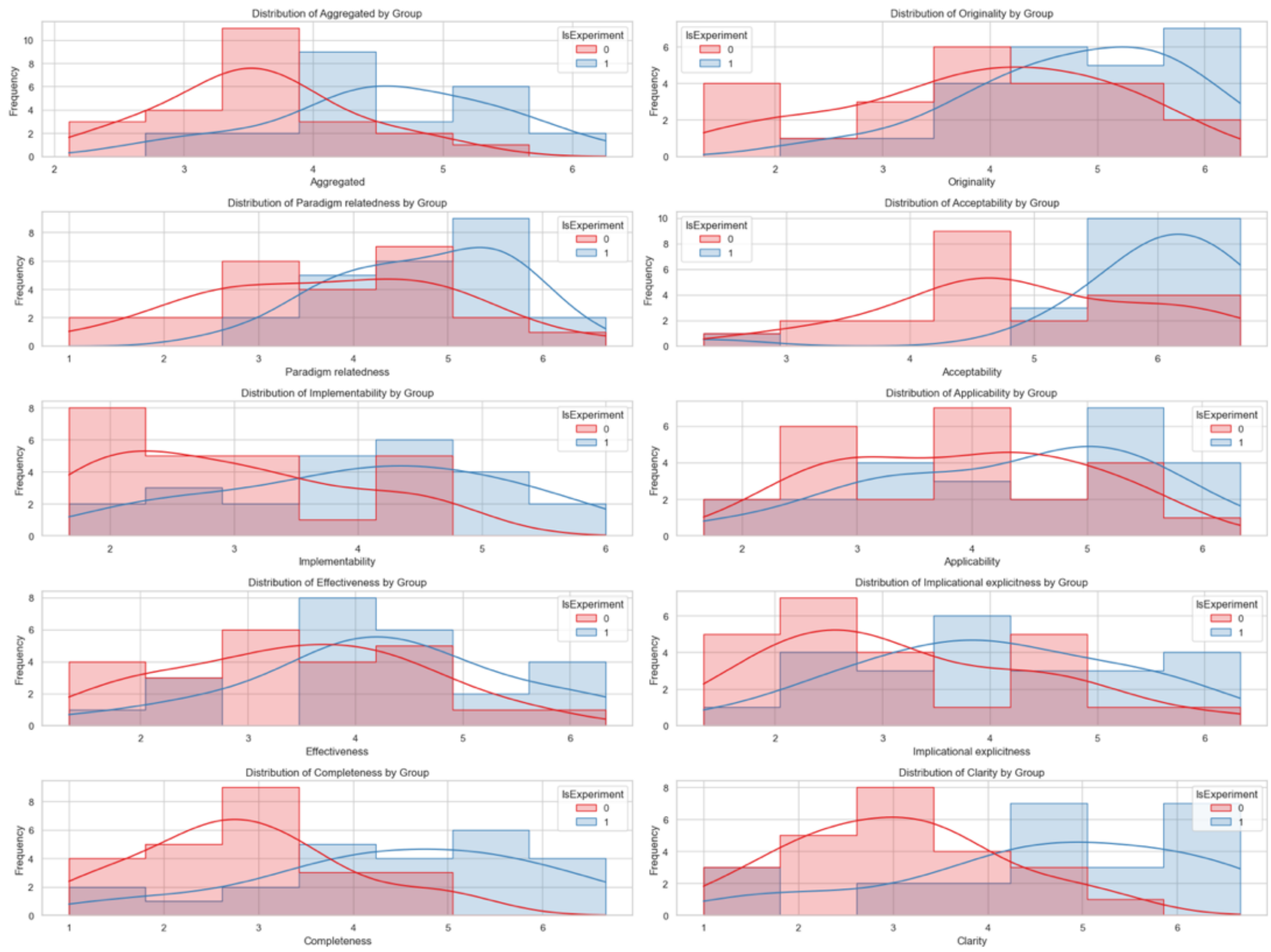

*Figure 2: Overlaid histograms with kernel density estimates*

The combined dataset reveals that the experimental group, assisted by GenAI, achieved higher scores in nearly all evaluation criteria, except for the applicability of the proposed ideas. The histograms indicate a marked rightward shift in the aggregate results for the experimental group, signifying a significant performance enhancement attributable to AI integration.

This trend was also observed in other key dimensions, such as originality and paradigm relevance. However, the analysis of feasibility presented a more nuanced picture. While the control group had slightly lower overall scores, they exhibited a marginally higher peak frequency in feasibility, necessitating a deeper analysis to draw conclusions. Nonetheless, the experimental group exhibited significantly higher scores in effectiveness, implicit clarity, completeness, and clarity. These dimensions suggest that ideas developed with the assistance of AI are more effective, clearly articulated, and comprehensive.

Figure 3 provides a detailed examination of the mean values across the individual evaluation categories, underscoring the superior performance of the experimental group. In every measured variable—aggregate score (Experimental (E): 4.583, Control (C): 3.546), originality (E: 4.819, C: 3.889), paradigm relevance (E: 4.75, C: 3.75), acceptance (E: 5.931, C: 4.903), feasibility (E: 4.028, C: 2.944), applicability (E: 4.333, C: 3.875), effectiveness (E: 4.236, C: 3.444), implicit clarity (E: 4, C: 3.25), completeness (E: 4.5, C: 2.833), and clarity (E: 4.653,

C: 3.028)—the experimental group scored significantly higher. On average, the experimental group's measured dimensions were 30% better than those of the control group (minimum: 21% for acceptance, maximum: 59% for completeness). These findings are also consistent with the medians of these variables.

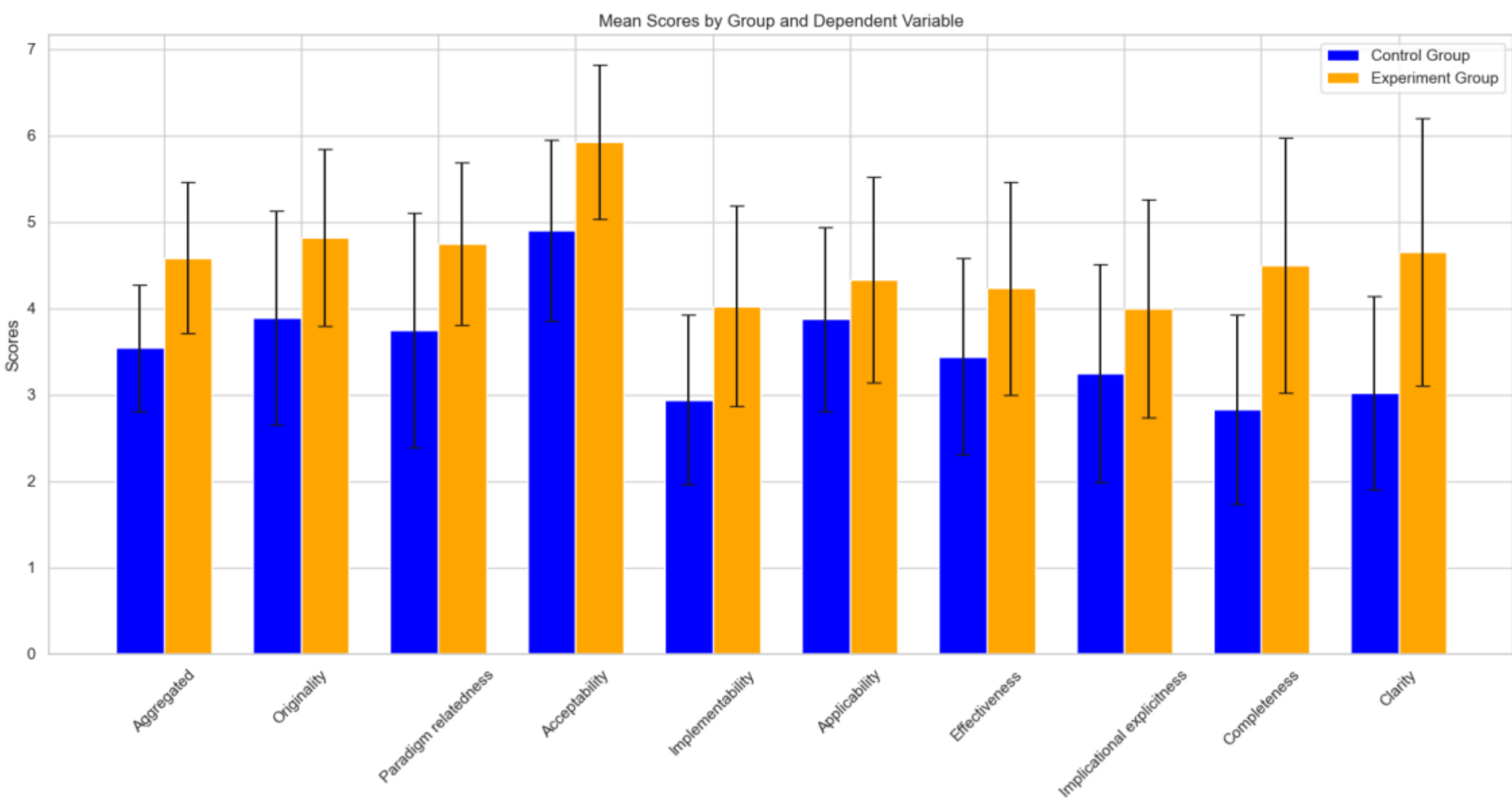

*Figure 3: Histogram of the average values per evaluation category*

When categorizing the ideas using the Abernathy and Clark innovation matrix, a similar trend emerged. The control group primarily developed niche innovations, while the experimental group, utilizing GenAI, tended toward more revolutionary innovations. For the analysis of the innovation types, only the ideas from Task 1 were considered, as Task 2's nature had an already defined problem definition and could significantly skew the results.

A significant trend was observed in the time measurement data, indicating that the experimental group developed solutions more efficiently across both tasks. In the first task, the experimental group was on average 6% faster than the control group (E: 39 min; C: 42 min), while in the second task, the difference was much more pronounced, with the experimental group completing it nearly 30% faster (E: 27 min; C: 38 min). It is important to note that although these were two different tasks, and the experimental group still needed to familiarize themselves with the GenAI-augmented ideation tool.

A post-study survey measured participant satisfaction with LLMs using a 10-point Likert scale. The control group, which ideated without GenAI support, had an average satisfaction score of 7.48. In contrast, the experimental group, supported by LLMs, reported a significantly higher average score of 9.22, reflecting a 23.3% increase in satisfaction.

To gain a deeper understanding, inductive statistical methods were employed. Initially, the Shapiro-Wilk test was conducted to assess the assumption of normality. All variables except for feasibility in the control group exhibited a normal distribution. Subsequently, Levene's tests were used to confirm the homogeneity of variances between groups, allowing for the application of t-tests to examine the statistical significance of differences.

*Table 2:* *T-test results of the evaluation criteria*

| Category | t-test | p-Value |
| --- | --- | --- |
| Aggregated | -4.450938 | 0.000056* |
| Originality | -2.840212 | 0.00678* |
| Paradigm Relatedness | -2.970548 | 0.004946* |
| Acceptability | -3.6546 | 0.000673* |
| Implementability | -3.495569 | 0.001079* |
| Applicability | -1.403744 | 0.167191 |
| Effectivity | -2.308438 | 0.02555* |
| Implicit Explicitness | -2.053805 | 0.045704* |
| Completeness | -4.427445 | 0.000066* |
| Clarity | -4.163475 | 0.000153* |

Table 2 presents the results of independent t-tests for various evaluation variables. A significant performance advantage for the experimental group in overall ideation performance was confirmed by a t-statistic of -4.450938 and a remarkably low p-value of 0.000056. Significant improvements were also observed in the originality and paradigm-related variables, with t-values of -2.840212 and -2.970548, and p-values of 0.00678 and 0.004946, respectively.

*Table 3: Results of the linear regression*

| Category | t-test | p-Value | R-squared |
| --- | --- | --- | --- |
| Aggregated | 1.037037 | 0.000054 | 0.301027 |
| Originality | 0.930556 | 0.006694 | 0.149201 |
| Paradigm Relatedness | 1 | 0.004713 | 0.160954 |
| Acceptability | 1.027778 | 0.000659 | 0.225016 |
| Implementability | 1.083333 | 0.001059 | 0.20988 |
| Applicability | 0.458333 | 0.16711 | 0.041077 |
| Effectivity | 0.791667 | 0.025521 | 0.103819 |
| Implicit Explicitness | 0.75 | 0.045704 | 0.083996 |
| Completeness | 1.666667 | 0.000058 | 0.298805 |
| Clarity | 1.625 | 0.000136 | 0.273698 |

The linear regression analyses summarized in Table 3 highlight the impact of GenAI support across various evaluation categories. The overall ideation score, with a coefficient of 1.037 and a p-value of 0.000054 ($R^2 = 0.3$), indicates a significant enhancement in idea quality. Originality also shows improvement (coefficient 0.93, $R^2 = 0.15$), though with some influence from external factors. Paradigm-relatedness is positively affected (coefficient 1, $R^2 = 0.16$). Significant gains are observed in acceptability (coefficient 1.03) and feasibility (coefficient 1.08), while applicability demonstrates a weaker effect (coefficient 0.458, non-significant p-value). Effectiveness and implicit clarity have meaningful coefficients but moderate $R^2$ values. Notably, completeness (coefficient 1.67) and clarity (coefficient 1.625) display strong improvements, as evidenced by high $R^2$ values (~30%) and low p-values, underscoring substantial benefits from GenAI tools in these areas.

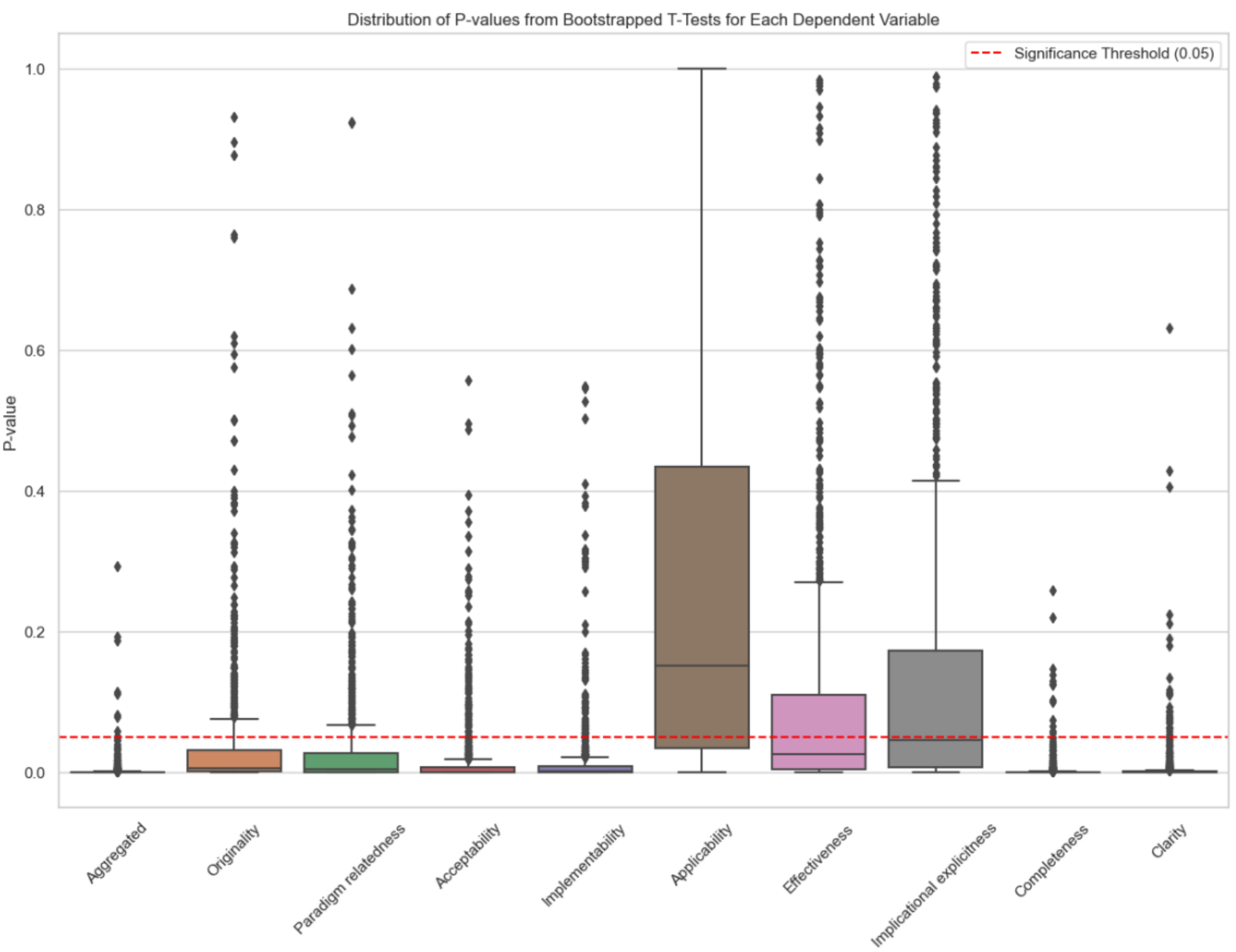

*Figure 4: Sensitivity Analysis Results - 1000 Bootstrap Variations*

A bootstrap procedure with 1000 resampled iterations adds robustness to the analysis by assessing the consistency of significance tests across a range of simulated and varied samples. The box plot in Figure 4 shows a strong and stable correlation between AI usage and improved ideation metrics like evaluation criteria, originality, and clarity, all with mean p-values well below 0.05. However, the relationship with applicability, effectiveness, and implicit clarity was less consistent, with mean p-values often above the significance threshold.

Cohen's d was also used to measure the standardized effect size for various dimensions of ideation. As shown in Figure 5, the effect sizes for variables such as aggregated evaluation criteria, and consequently idea quality, originality, completeness, and clarity, yielded Cohen's d values above 0.8, indicating large effect sizes and a significant AI effect. In contrast, similar to the bootstrap procedure, the effect sizes for applicability, effectiveness, and implicit clarity were smaller. Medium effect sizes for paradigm-relatedness and acceptability suggest a significant, though not transformative, role of AI.

The analysis focused on Task 1 to avoid bias from Task 2's predefined problem (autonomous driving) in the innovation categories of the Abernathy and Clark matrix. The point-biserial correlation coefficient was used to assess the relationship between GenAI-assisted ideation and innovation types. Results indicated moderate positive correlations in GenAI-assisted groups for regular ($r = 0.209$, $p = 0.328$) and revolutionary innovations ($r = 0.275$, $p = 0.193$), and slight negative correlations in the control group for niche ($r = -0.251$, $p = 0.237$) and architectural innovations ($r = -0.103$, $p = 0.633$). However, none of these correlations were statistically significant ($p > 0.05$), highlighting trends without a definitive link between GenAI and innovation type. Further research with larger samples is recommended to clarify GenAI's impact on ideation.

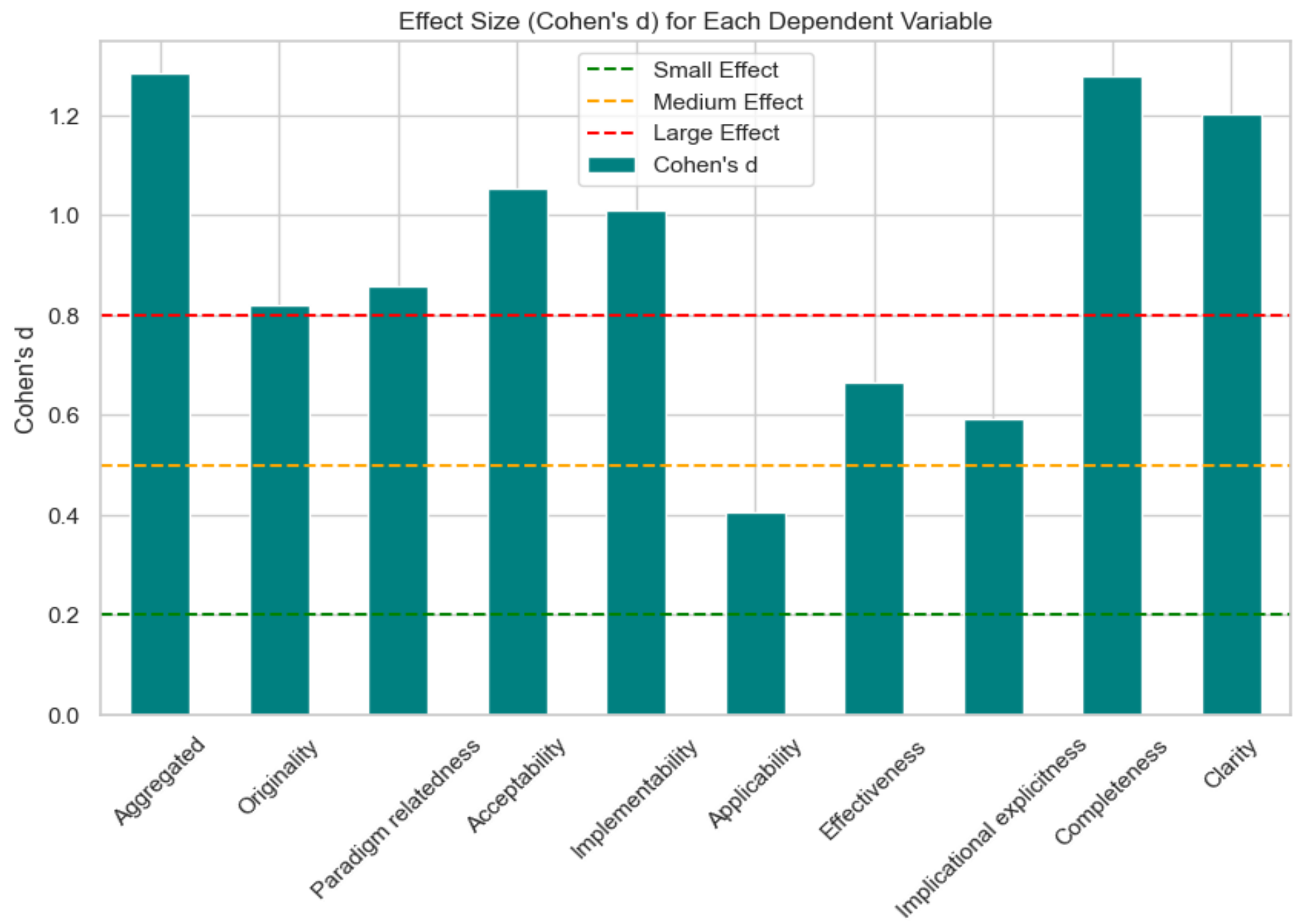

*Figure 5: Effect Size Results According to Cohen's d*

The t-test for time measurement showed that the experimental group reached a higher-quality problem solution faster. A t-value of 3.92 with a very low p-value of 0.00029 indicates that the use of GenAI-augmented ideation tools in the experimental group led to a significant reduction in the time required for ideation tasks compared to the traditional method in the control group.

Regarding the impact of GenAI on ideation and innovation team performance, the analysis of the results provides compelling evidence for several hypotheses. First, the integration of LLMs significantly enhanced knowledge spillover, leading to higher-quality ideas in terms of acceptability, feasibility, completeness, and clarity (H1). Additionally, LLMs accelerated the ideation process, improving time efficiency by 6% for Task 1 and 30% for Task 2, with a t-value of 3.92, supporting H2. The GenAI-augmented group not only generated better ideas but also increased their idea output per unit of time, confirming H3. The diversity in idea novelty, feasibility, and relevance was notably higher in the GenAI-augmented group, supporting H4, although further investigation is needed for specific diversity dimensions. The trend towards generating more revolutionary ideas indicates a shift in ideation, partially confirming H5, with larger datasets required for full validation. Lastly, H6 was confirmed by the higher satisfaction and engagement levels among participants using LLMs.

# 6 Discussion and Implications

The analysis shows a rightward shift in the performance metrics of the experimental group, indicating improved outcomes in idea quality. While the control group showed slightly lower overall scores, they had higher peaks in feasibility, which warrants further exploration. The experimental group demonstrated significantly higher effectiveness, clarity, completeness, and overall quality of ideas, highlighting GenAI's positive impact on ideation. Inductive analysis confirmed that the experimental group performed significantly better in overall idea generation, with notable improvements in originality, paradigm relevance, acceptance, and feasibility. The study also found that the experimental group was more likely to produce revolutionary innovations. This shows that AI augmented leads to ideas with a higher disruptive potential, although a more detailed examination is needed. A satisfaction survey revealed a 23.3% higher satisfaction rate in the experimental group, suggesting that AI integration can significantly boost participant satisfaction and engagement in the ideation process. Time measurements indicated clearly that the experimental group reached high-quality solutions more quickly, supporting hypotheses of improved knowledge spillover, accelerated ideation, and increased efficiency through GenAI. AI systems can recognize problems, opportunities, and threats beyond local search routines and knowledge domains. [10] Examples include analyzing online discussion forums to extract customer needs [94] and solving complex scientific problems like protein folding. [114]

In the context of GenAI-augmented ideation and its impact on creativity and performance within innovation teams, this research contributes to existing theories by showing that LLMs can significantly enhance team creativity through their extensive knowledge base and sophisticated tools for manipulating that knowledge. The improvement is driven by the role of inspirational stimuli in fostering creativity, resulting in original and actionable solutions that challenge established paradigms. Additionally, GenAI-augmented ideation addresses common challenges in group brainstorming, such as the reluctance to share ideas and production blockages, thereby enhancing collaboration and increasing idea diversity. The findings indicate that GenAI-augmented CSS improve both the efficiency and quality of ideation compared to traditional methods and archive a higher satisfaction among ideation teams.

The research enhances our understanding of the Knowledge Spillover Theory of Entrepreneurship. LLMs are instrumental in translating tacit knowledge into explicit forms, thereby reducing knowledge filters and enhancing absorptive capacity—key elements of the KSTE. This transformation in knowledge dynamics has the potential to enrich individuals and innovation teams and exert a substantial influence on the field of entrepreneurship and innovation management. The study posits that GenAI-augmented CSS mitigate knowledge filters by increasing knowledge accessibility and broadening absorptive capacity through contextual expansion. On an individual level, this could lead to the generation of a higher number of high-quality ideas, offering competitive advantages to both established and emerging firms. From a macroeconomic standpoint, AI is positioned as a transformative force in entrepreneurship and innovation, essential for promoting knowledge spillovers and facilitating the dissemination and application of knowledge in economic activities. This expanded perspective may impact elements of the Zeitgeist and Communication theory of innovation, which holds that major breakthroughs occur primarily in areas central to collective intellectual efforts.

The data analysis supports the application of KSTE, confirming that entrepreneurial opportunities arise from the dissemination and absorption of knowledge. The impact of AI is evident, as GenAI acts as a catalyst in the knowledge spillover process, particularly through LLMs that accelerate knowledge exchange and enhance idea generation quality. The impact of GenAI is particularly significant during the ideation phase of innovation, where it enhances the idea generation, potentially accelerating the development of novel and valuable concepts. This impact is anticipated to extend to subsequent stages of innovation, including prototyping and implementation, prompting the need for theoretical adjustments in innovation process models. Ultimately, the incorporation of AI into ideation and creative processes not only augments the ideation environment but also signifies a paradigm shift in the understanding and management of knowledge spillovers and innovation processes.

In the field of innovation management, particularly during the ideation phase, the concept of the 'Jagged Frontier' plays a central role [149]. This term highlights the uneven capabilities of AI, as AI excels in tasks typically challenging for humans, while surprisingly struggling with seemingly simpler tasks. The concept is crucial for identifying situations where AI may not be effective. Despite these challenges, the boundaries of AI capabilities are continuously expanding, as various studies in different fields demonstrate. In legal and consulting fields, AI support shows significant improvements in task completion, speed, and quality [149], [150]. When GenAI is employed by consultants, task completion rates increased by 12.2%, processing speed by 25.1%, and quality by 40%. Similar effects are observed in programming, where the use of GenAI, particularly the AI tool GitHub Copilot, led to a remarkable productivity increase of 55.8% [151]. Several recent studies investigating the impact of GenAI on ideation, either by isolating AI-generated ideas or without assessing efficiency, have demonstrated significant enhancements in novelty and creativity when AI tools are utilized in the ideation process [5], [152], [153], [154].

This study reinforces and extends these earlier observations by focusing on hybrid AI human innovation teams and suggests that these findings are particularly relevant for innovation management, especially in the ideation phase. It was shown that students, with GenAI support, required in average 27.5% less time to generate ideas of 29.24% higher quality. These results correspond with the general impact of GenAI on various professions. The significant reduction in time required and the simultaneous increase in idea quality using GenAI tools underscore the potential of GenAI to substantially optimize innovation processes. This has far-reaching implications for companies and organizations striving to enhance their innovation capabilities.

The influence of GenAI on enhancing idea quality is profound across various domains of innovation management and entrepreneurship. Ideas serve as the bedrock of innovation, and in a rapidly evolving, highly competitive technological landscape, the imperative to "innovate or die" has never been more pertinent. In this context, continuous generation of novel and creative solutions is essential for maintaining competitiveness and addressing the swiftly changing demands of the market. As observed, GenAI tools significantly elevate the novelty (originality, paradigm relatedness), feasibility (acceptability, implementability), relevance (applicability, effectiveness), and specificity (implicational explicitness, completeness, clarity) of ideas.

GenAI algorithms excel at analyzing and integrating vast datasets to uncover unique patterns and opportunities. LLMs contribute to the development of original ideas by processing extensive data in creative and innovative ways. These systems also simulate and predict outcomes, thereby bolstering the practicality and feasibility of ideas. GenAI facilitates the assessment of an idea's feasibility and potential effectiveness before resource allocation, promoting a more efficient innovation strategy and increasing the likelihood of success for new projects and products. The application of GenAI in idea generation and evaluation thus plays a crucial role in innovation management, offering companies a competitive edge through improved idea completeness and clarity.

GenAI technologies offer extensive background information, highlighting aspects often overlooked by human ideators. By simulating multiple perspectives, GenAI aids in developing more comprehensive concepts and translating complex ideas into clearer, more accessible formats. This study demonstrates how GenAI enhances the completeness and clarity of ideas, innovations, and business strategies, leading to more effective communication within teams and with external stakeholders, such as venture capitalists. Improved communication strengthens team alignment and boosts capital acquisition opportunities, giving startups and organizations using GenAI a competitive edge. For instance, in the pharmaceutical industry, AI-driven data analysis and GenAI help discover innovative and feasible drug formulations, expanding entrepreneurial opportunities and improving organizational outcomes [155].

Additionally, GenAI significantly improves the time efficiency of idea generation, reducing brainstorming and ideation time by 25-40%. This allows teams to focus more on refining ideas, testing, or integrating diverse perspectives, ultimately enhancing productivity and effectiveness. GenAI-driven idea generation increases the quality of ideas per unit of time, leading to productivity gains and influencing team dynamics and resource allocation. Integrating GenAI into creative processes requires adjustments in management, resource evaluation, and skill development, positioning GenAI as a strategic asset in a competitive digital landscape—a trend evident in startups and venture capital allocation [156].

The potential of GenAI to drive revolutionary innovations is substantial. GenAI-powered tools can transcend conventional thinking, fostering groundbreaking innovations. While further research is needed, current evidence suggests that developing GenAI for revolutionary thinking could trigger transformative changes across various fields, similar to the impact of the Internet revolution on society, politics, and the economy.

The study highlights the importance of integrating generative AI tools, especially for ideation, to enhance creative processes in innovation. Research suggests embedding these tools throughout the innovation process, focusing first on ideation to expand opportunities and improve collaboration. To manage the increased idea output, organizations need strong evaluation frameworks, guided by a clear vision and strategic goals. AI models should be tailored to organizational needs, with a blend of AI and traditional knowledge management. Training teams to work effectively with AI is crucial for better results, and regular audits are needed to correct AI biases. Despite AI's advancements, human creativity and input remain vital due to AI's limitations.

# 7 Limitations & Future Outlook

This study has several limitations that must be considered when interpreting its findings on the impact of GenAI on ideation within innovation teams. One challenge is isolating AI's specific effects from other factors, such as team dynamics and individual creativity. Although participants were randomized into similar groups, pre-existing knowledge could have influenced outcomes, complicating the assessment of GenAI's true impact and the knowledge spillover facilitated by LLMs. The focus on explicit knowledge gained through GenAI may not fully capture the teams' nuanced understanding or application, particularly concerning implicit knowledge. The study's generalizability is also limited by the homogeneous participant pool, which may not reflect the diversity of real-world innovation teams across various industries. Furthermore, the study's emphasis on specific innovation tasks in healthcare and the automotive industry limits its broader applicability.

The potential cognitive load and information overload from AI tools were not thoroughly explored, despite their possible significant impact on ideation. Ethical considerations, such as AI-induced bias and intellectual property implications, were also underexamined. The lack of long-term observations to assess AI's lasting impact on ideation and team dynamics, along with the learning curve associated with new AI tools, further constrains the study's conclusions. Although the study analyzed the qualitative outputs of experimental and control groups, it does not fully explain the role of knowledge spillover from LLMs in the experimental group's superior performance, particularly given the subjective nature of idea evaluation. The use of a heuristic evaluation method according to Dean et al. (2006) and a diverse jury was noted, but the inherent subjectivity in assessing unrealized ideas remains a limitation.

Moreover, potential biases in AI-generated suggestions could limit the generation of radical innovations due to AI's reliance on existing data, although recent studies suggest AI can contribute to creative ideas [4]. This study does not fully address AI's potential to foster dependency among team members, potentially inhibiting their ability to innovate without technological support.

Future research should delve deeper into the Knowledge Spillover Theory of Entrepreneurship, specifically examining GenAI's impact on creativity and innovation. Key areas include the role of LLMs in knowledge dissemination, distinguishing between user and AI contributions, and optimizing knowledge spillover. Further studies should assess AI's effect on innovation ecosystems and team dynamics, using surveys or interviews for more insight. Evaluating AI-assisted ideation tools, particularly in idea quality, cognitive load reduction, and value compared to traditional methods, is crucial. The shift towards smaller, efficient, and open-source LLMs, which now rival commercial models and especially the much higher context window of future AI, also merits further exploration.

# 8 Conclusion

In this empirical study, the impact of AI on idea generation and the performance of innovation teams was analyzed. Participants, consisting of students, were divided into two groups: a control group, which engaged in ideation without AI support, and an experimental group that utilized a custom-developed AI-assisted ideation tool. Teams of three participants were tasked with solving two distinct problems. Upon completion, the ideas generated by both groups were evaluated by a panel of experts in academia, entrepreneurship, and innovation management. This panel assessed the ideas based on the dimensions of idea quality proposed by Dean et al. (2006) [20], without being informed of the group affiliations. Following this assessment, the ideas were categorized using Abernathy and Clark's (1985) innovation matrix [82].

The study's findings clearly demonstrate that teams utilizing AI support generated higher-quality ideas in a shorter amount of time. Across all dimensions of idea quality, except for applicability, the AI-assisted teams outperformed the control group. AI usage enhanced knowledge sharing (H1) accelerated the ideation process, and increased efficiency (H2, H3). Additionally, the diversity of ideas, particularly in terms of novelty and feasibility, was greater in the AI-supported group (H4). The study also suggests that AI has the potential to lead to breakthrough innovations (H5). AI-assisted teams exhibited higher satisfaction and engagement (H6), underscoring the critical role of AI in enhancing creativity and efficiency during the ideation phase. These results imply that AI has significant potential to influence entrepreneurial creativity and improve the exploitation of entrepreneurial opportunities. The effects observed in this study align with findings from other research on the impact of generative AI in fields such as programming, legal work, and consulting. These insights contribute to a broader understanding of the areas in which AI support yields positive outcomes.

The primary objective of this research was to analyze and quantify the influence of artificial intelligence on the dynamics of innovation teams during the ideation phase. Through the examination of empirical data from a controlled field experiment, this study aimed to provide concrete insights into the role of AI in shaping team creativity and the resulting innovation outcomes. This objective was achieved by identifying that LLMs have a profound impact on creativity and knowledge dynamics, thereby directly and indirectly influencing key aspects of the Knowledge Spillover Theory of Entrepreneurship.

In summary, the study's findings underscore the significant role of AI, particularly GenAI, in fostering creativity and entrepreneurial opportunities. By optimizing the ideation process, promoting a diverse and innovative range of ideas, enhancing user engagement, and strengthening knowledge spillover, AI emerges as a central component of the modern entrepreneurial paradigm. Within the context of the Knowledge Spillover Theory of Entrepreneurship, these findings also reinforce the understanding that advancements in artificial intelligence are driving shifts in innovation and business dynamics. Furthermore, the study supports the thesis that the synergy between knowledge and technological progress serves as a catalyst for future entrepreneurial opportunities and, consequently, economic growth.

# Appendix

## Appendix 1: AI augmented ideation tool

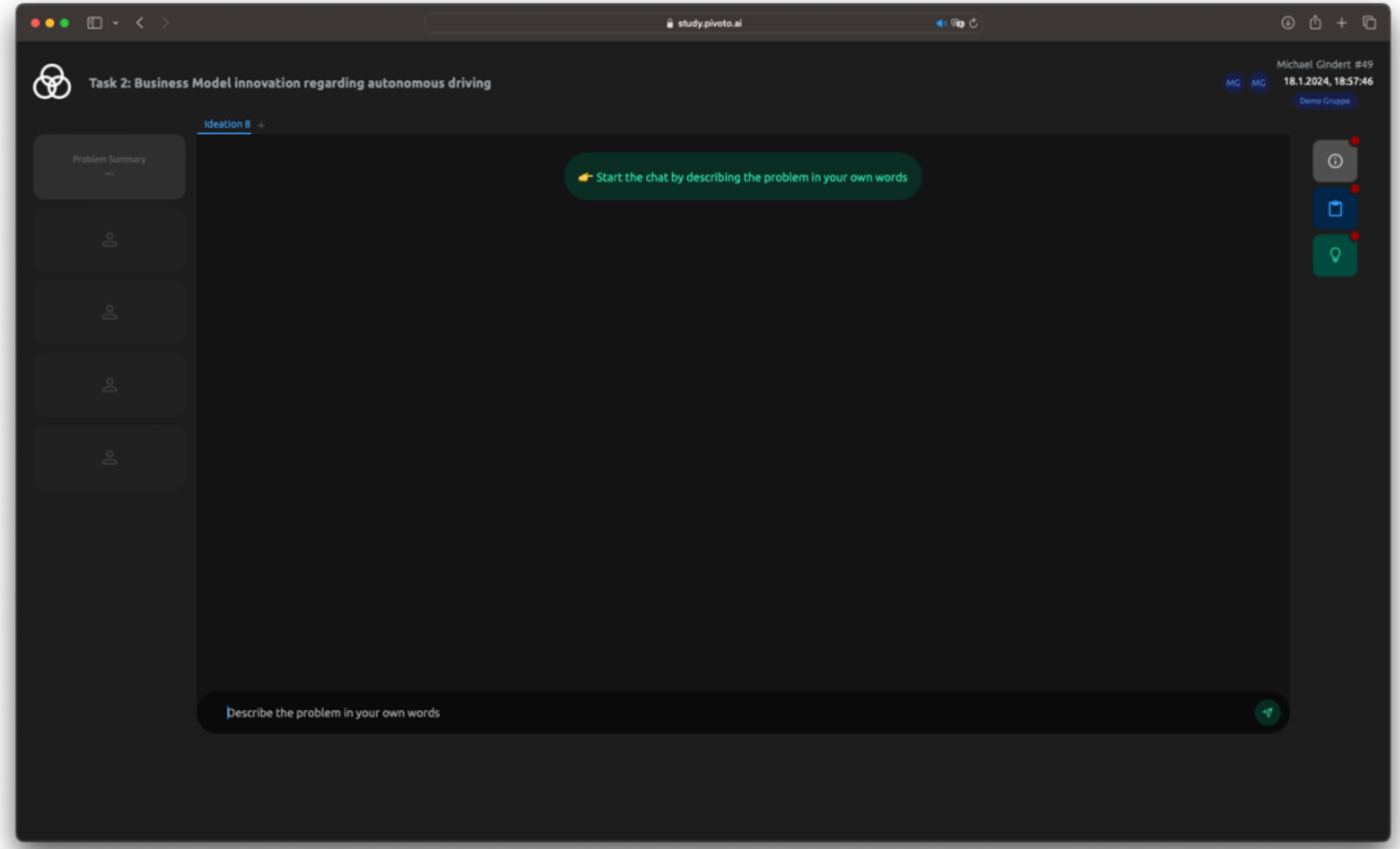

*Screenshot 1: Overview of the tool*

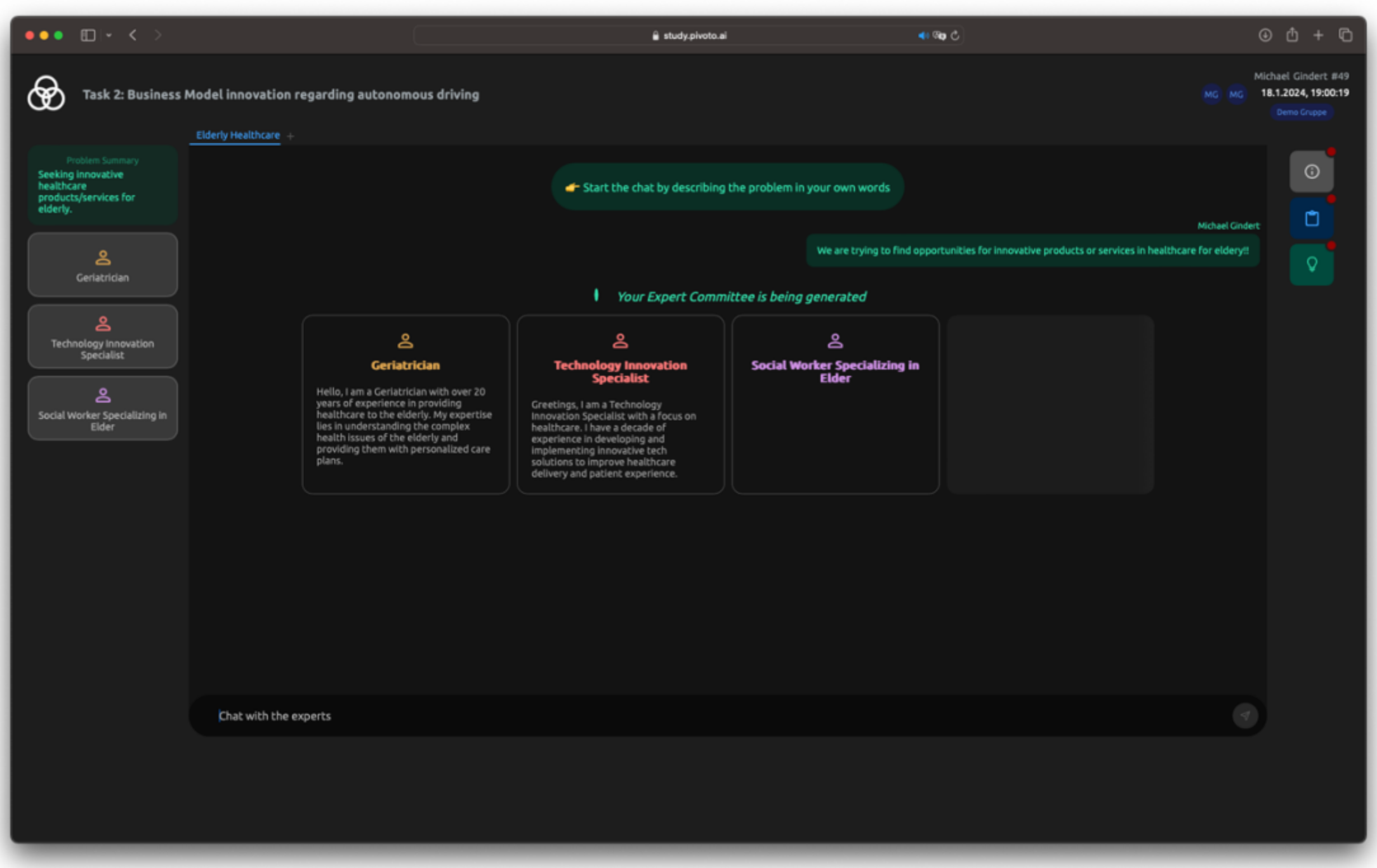

*Screenshot 2: AI persona generation based on problem definition*

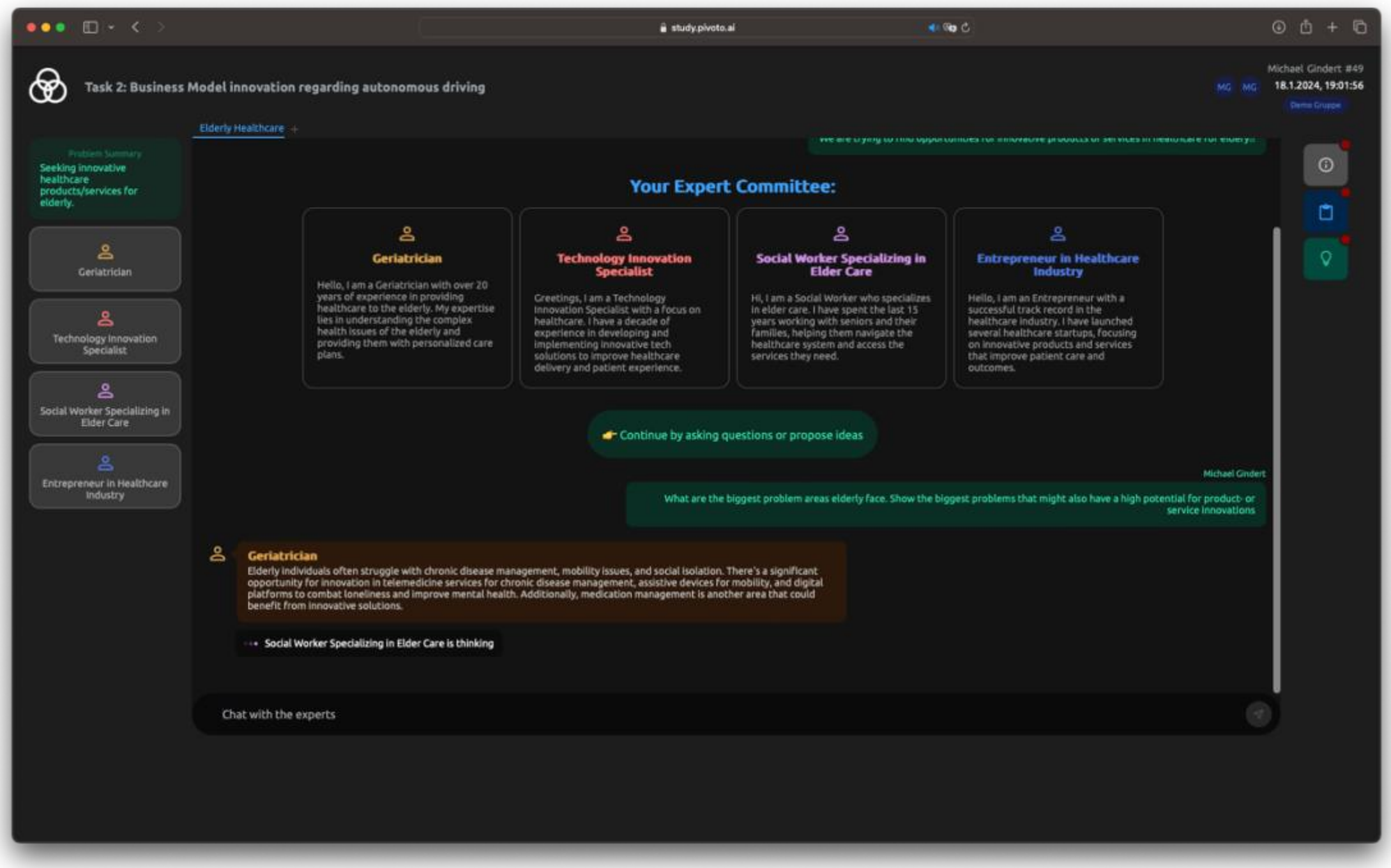

*Screenshot 3: AI personas contribute from persona perspective*

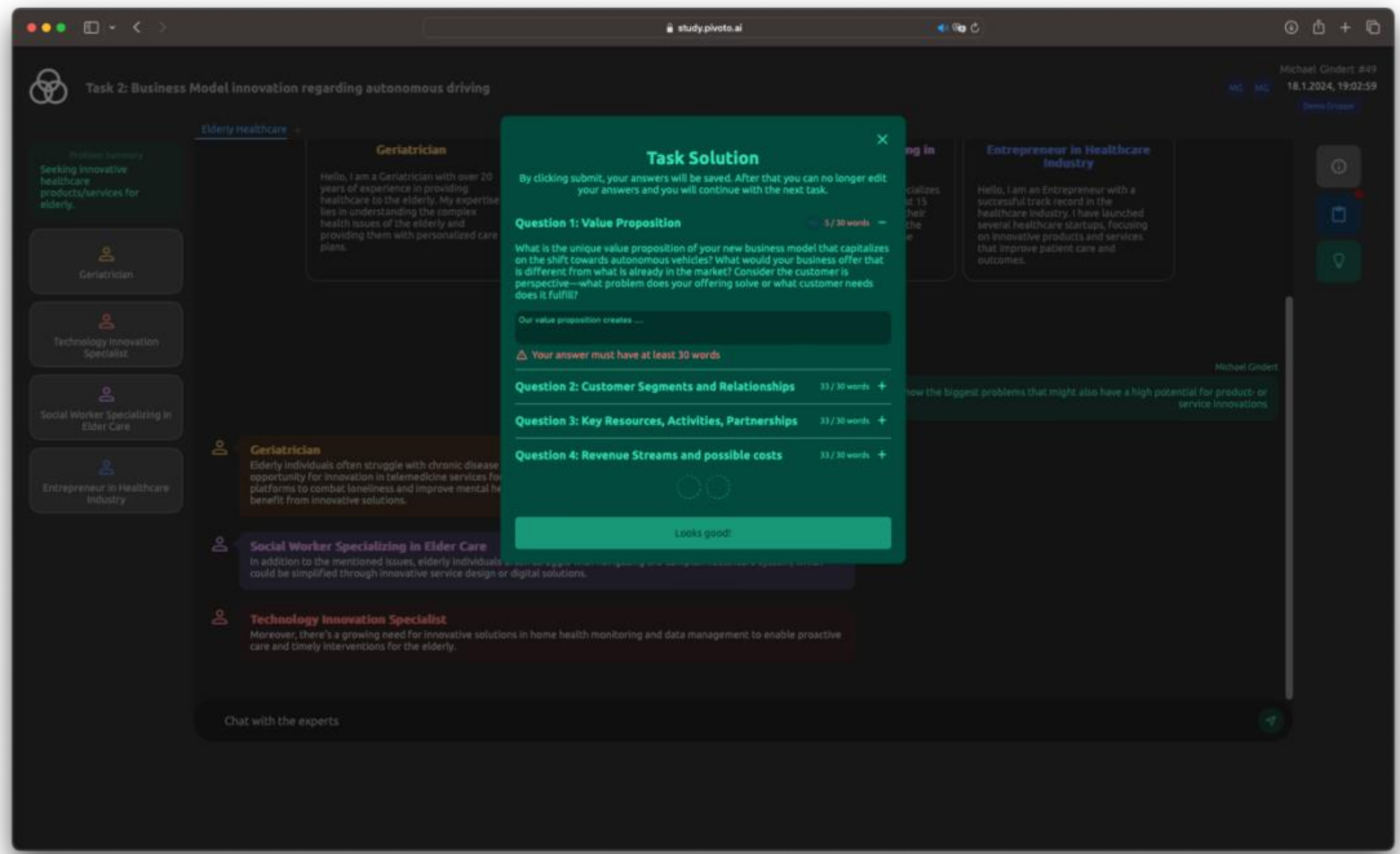

*Screenshot 4: Task solution input*

## Acknowledgements

We would like to express our deepest gratitude to Professor Dr. Michael Dowling and the Chair of Innovation Management for their invaluable guidance and support. Special thanks to Marc Emanuel Alexander Otto for his unwavering assistance, and to our parents, friends, and everyone who contributed to this work.

Your encouragement and help made this achievement possible.

## Declaration of generative AI in scientific writing

During the preparation of this work the author(s) used GPT-4 in order to increase readability. After using this tool/service, the author(s) reviewed and edited the content as needed and take(s) full responsibility for the content of the published article.